\documentclass{article}
\usepackage{arxiv}

\usepackage{tabularx} 
\usepackage{amsmath}  
\usepackage{amssymb} 
\usepackage{graphicx} 
\usepackage{wrapfig}
\usepackage{subfig} 
\usepackage{tikz} 
\usetikzlibrary{arrows} 

\usepackage[final]{hyperref} 
\usepackage{accents} 
\usepackage{pdfpages}
\usepackage{appendix}
\usepackage{placeins} 
\usepackage{mathtools} 
\usepackage{dsfont}  
\usepackage{algorithm}
\usepackage{algpseudocode}
\usepackage[numbers,sort&compress]{natbib}  
\hypersetup{
	colorlinks=true,       
	linkcolor=blue,        
	citecolor=blue,        
	filecolor=magenta,     
	urlcolor=blue         
}
\graphicspath{{../PhD/figures/}}
\newcommand{\diff}{{\rm{d}}}

\DeclareMathOperator*{\argmax}{\arg\!\max}
\usepackage{amsthm}
\theoremstyle{definition}

\begin{document}

\title{A Monte Carlo EM Algorithm for the Parameter Estimation of Aggregated Hawkes Processes}




\author{
  Leigh~Shlomovich\\
  Department of Mathematics\\
  Imperial College London\\
  South Kensington, London, SW7 2AZ \\
  \texttt{leigh.shlomovich14@imperial.ac.uk} \\
   \And
  Edward~A.~K.~Cohen\\
  Department of Mathematics\\
  Imperial College London\\
  South Kensington, London, SW7 2AZ \\
  \texttt{e.cohen@imperial.ac.uk} \\
  \And
  Niall~Adams\\
  Department of Mathematics\\
  Imperial College London\\
  South Kensington, London, SW7 2AZ \\
  \texttt{n.adams@imperial.ac.uk} \\
  \And
  Lekha~Patel\\
  Department of Mathematics\\
  Imperial College London\\
  South Kensington, London, SW7 2AZ \\
  \texttt{lekha.patel11@imperial.ac.uk} \\
}



\maketitle

\begin{abstract}

A key difficulty that arises from real event data is imprecision in the recording of event time-stamps. In many cases, retaining event times with a high precision is expensive due to the sheer volume of activity. Combined with practical limits on the accuracy of measurements, aggregated data is common. 
In order to use point processes to model such event data, tools for handling parameter estimation are essential.
Here we consider parameter estimation of the Hawkes process, a type of self-exciting point process that  
has found application in the modeling of financial stock markets, earthquakes and social media cascades. We develop a novel optimization approach to parameter estimation of aggregated Hawkes processes using a Monte Carlo Expectation-Maximization (MC-EM) algorithm.
Through a detailed simulation study, we demonstrate that existing methods are capable of producing severely biased and highly variable parameter estimates and that our novel MC-EM method significantly outperforms them in all studied circumstances. These results highlight the importance of correct handling of aggregated data.

\end{abstract}


%


\section{Introduction}



Point processes are extensively used to model event data and have found wide applications in many fields including seismology \cite{ogata_seismicity_1999} and cyber-security \cite{price-williams_nonparametric_2019}. One representation of a point process is via the counting process $\{N(t), \, t \in \mathbb{R}\}$, where $N(t)$ denotes the number of events up to time $t$, and $N(0)=0$. Due to limited recording capabilities and storage capacities, retaining event times with a high precision is expensive and often infeasible. Therefore, in much real-world data, it is common to instead observe a times series of the aggregated process,
$$
N_t = N\left(\Delta (t+1)\right) - N \left( t \Delta \right),
$$
for some $\Delta > 0$, which we refer to as the \emph{bin width}. This aggregated process may arise from a predetermined binning of the data into counts, or equivalently from the rounding of event times. In the context of network traffic data, for example, the resolution of the recorded times can be anywhere from milliseconds to seconds (as is the case with the Los Alamos National Laboratory (LANL) NetFlow data \cite{turcotte_unified_2018}), or even coarser. In this setting the value of this aggregated process at each time point is the number of events with that rounded time-stamp. 

Intuitively, when aggregating data we lose information and essentially `blur' our view of the continuous time point process, making it potentially problematic to apply methods which assume a continuous time framework. Thus, the problem we consider here is to infer upon the underlying continuous process from the observed aggregated data.


The Hawkes process is a type of `self-exciting' process which provides us with a model for contagious event data. Their flexibility and real-world relevancy has resulted in a host of applications. In the case of financial data for example, this allows propagation of stock crashes and surges to be modeled \cite{bacry_non-parametric_2012, bowsher_modelling_2007, filimonov_quantifying_2012, fonseca_hawkes_2014, embrechts_multivariate_2011}. Propagation of social media events has also been modeled using Hawkes processes, in particular `twitter cascades' are considered in \cite{rizoiu_tutorial_2017,kobayashi_tideh:_2016}. Further applications include the modeling of civilian deaths due to insurgent activity in Iraq \cite{lewis_nonparametric_2011}, and predicting origin times and magnitudes of earthquakes \cite{ogata_statistical_1988}. 

Formally, the Hawkes process is a class of stochastic process with the defining property that 
\begin{align*}
\Pr\{ \diff N(t) = 1 | N(s) \; (s \leq t) \} &= \lambda^\ast(t) \diff t + o(\diff t), \nonumber \\
\Pr\{ \diff N(t) > 1 | N(s) \; (s \leq t) \} &= o(\diff t),
\end{align*}
where $\diff N(t) = N(t + \diff t) - N(t)$. It is characterized via its conditional intensity function (CIF) $\lambda^\ast(t)$, defined as
$$
\lambda^\ast (t) = \nu + \int_{-\infty}^{t} g(t-u){\mathrm{d}}N(u),
$$
where $\nu$ is called the background intensity and $g(u)$ is the excitation kernel. This means the intensity at an arbitrary time-point is dependent on the history of the process, producing self-exciting behavior. Depending on the kernel $g(u)$, the excitation may be quite local, or have longer term effects \cite{hawkes_spectra_1971}. 

We can also consider a Hawkes process as a branching process of time-stamped events. From this viewpoint, formalised in \cite{hawkes_cluster_1974}, events can be seen to arrive either via {\it{immigration}} or {\it{birth}}. That is, an event can be triggered by the background intensity rate $\nu$, 
in which case the event is seen as an immigrant. Alternatively, an event which is cause by self-excitation can be considered a descendant, referred to as being generated `endogenously'. Unlike a homogeneous Poisson point process, where events happen independently and at a constant rate, self-exciting processes are such that there is a higher likelihood of events happening in the near future of an arbitrary event and this is due to endogenous triggering \cite{rizoiu_tutorial_2017}. In this way we can consider the branching ratio of a Hawkes process, defined as 
\begin{align}\label{eqn:branching_ratio}
0 < \gamma := \int_{0}^{\infty} g(u) \diff u < 1.
\end{align}
The inequality above ensures that the process does not `explode', a case in which we have infinite events occurring in some finite time interval \cite{laub_hawkes_2015}. 

As introduced in Hawkes' original paper, exponential decay is a common choice for the excitation kernel due to the simplifications it provides for the theoretical derivations \cite{hawkes_spectra_1971,laub_hawkes_2015}. 
In this case we can write the excitation kernel as a sum of $L$ exponential decays,
\begin{align}\label{eqn:exp_kern_univariate}
g(u) = 
\begin{cases}
\sum_{l=1}^{L} \alpha_l \exp(-\beta_l u),&u > 0,\\
0, &{\rm{otherwise}}.
\end{cases}
\end{align}

Here, and as is most common, we let $L=1$ when considering the exponential kernel. In this case, the branching ratio defined in (\ref{eqn:branching_ratio}) becomes
\begin{align*}
\gamma = \int_{0}^{\infty} \alpha \exp(-\beta u) \diff u = \frac{\alpha}{\beta}.
\end{align*}
Given this model, we wish to estimate the parameter set $\Theta = \{\nu, \alpha, \beta\}$. Other kernels can be used, including a power-law function of form $g(u) = \alpha \beta {(1 + \beta u)^{-(1+c)}} \mathds{1}_{\mathbb{R}_+}(u)$, in which case $\Theta = \{ \nu, \alpha, \beta, c\}$. In the continuous time setting, parameter estimation for any of these kernels is straightforward. 

\subsection{Continuous Time Framework}

Typically, maximum likelihood estimation (MLE) is used to estimate parameters of a point process from a set of exact event times ${\mathcal{T} = \{t_1, \ldots, t_{N_{T}}\} \subset (0, T]}$. 
The CIF, or hazard function, of a point process is formally defined as  
\begin{equation}\label{eqn:cif_hazard}
\lambda^\ast (t) = \frac{f^\ast(t)}{1-F^\ast(t)},
\end{equation} 
where $f^\ast(t)$ and $F^\ast(t)$ are the conditional PDF and CDF, respectively, of the next arrival time, given the history of the process. From (\ref{eqn:cif_hazard}) we have that 
\begin{align}\label{eqn:cond_cdf_pdf}
F^\ast(t) &= 1 - \exp \left\{ - \int^t_{t_N} \lambda^\ast (u) \diff u\right\},\nonumber  \\
f^\ast(t) &= \lambda^\ast(t) \exp \left\{ - \int^t_{t_N} \lambda^\ast (u) \diff u\right\},
\end{align}
where $t_N$ denotes the last observed time prior to $t$. From (\ref{eqn:cond_cdf_pdf}), the joint likelihood of the univariate observations over the window $[0, T]$ is 
\begin{align*}
\mathcal{L}(\Theta ; \mathcal{T}) &= \prod_{j=1}^{N_T} f^\ast(t_j) (1-F^\ast(T)),\\
&= \left[ \prod_{j=1}^{N_T} \lambda^\ast(t_j) \right] \exp \left\{ - \int^{T}_{0} \lambda^\ast (u) \diff u\right\}. 
\end{align*}
Thus, from Proposition 7.2.III of \cite{daley_introduction_2003}, the log-likelihood is given by
\begin{align}\label{log_like_formula1}
\log \mathcal{L}(\Theta ; \mathcal{T}) &= \sum_{j=1}^{N_T} \log \lambda^\ast(t_j) - \int^{T}_{0} \lambda^\ast (u) \diff u.
\end{align}
If specifically considering a Hawkes process with exponential excitation kernel of form $\alpha \exp(-\beta t ),$ this log-likelihood can be simplified and expressed recursively as shown in \cite{laub_hawkes_2015}. In this paper we consider methods that are required when we instead observe a binned sequence of event counts. An alternative but equivalent representation of this is a discretization or rounding of the latent time-stamps, but here we will consider the observed data as the aggregation of $\mathcal{T}$ to bins. 



\subsection{Aggregated Data}



In the literature, the issue of aggregated data is handled in many ways, from uniformly redistributing events across the bin \cite{bowsher_modelling_2007}, to only retaining unique time-stamps and discarding the rest \cite{lorenzen_analysis_2012}. Here, we propose a novel Monte Carlo-Expectation Maximization (MC-EM) method and compare it to two existing approaches, evaluating the performance of parameter estimation for each. The methods compared are:

\begin{enumerate}
\item Approximating a binned Hawkes process as an integer-valued auto-regressive (INAR) process, a method developed in \cite{kirchner_hawkes_2016, kirchner_estimation_2017} and described in Section \ref{inar_section}.  
\item Formulating a binned log-likelihood which assumes a piecewise constant CIF within each interval as described in Section \ref{sec:binnedll}.
\item A novel MC-EM approach detailed in Section \ref{sec:MCEM}.
\end{enumerate}

There are other methods which have been covered in the literature, but are not considered here due to lack of applicability to this problem. As an example, a significant amount of the literature which aims to work with aggregated data considers binning the time-points such that the process contains at most one event per bin as in \cite{obral_simulation_2016}. This is inappropriate here as we do not have access to the latent event times and so cannot select an appropriate discretization level, $\Delta$. Likewise, some literature considers modeling binned behavior as a Bernoulli process \cite{brillinger_maximum_1988}. Again, this is invalid here as it fails to account for the number of events in a bin and thus will heavily bias results. There exist methods that handle missing data when we observe continuous time-points with gaps in the recording windows \cite{le_multivariate_2018}. That is, when the data considered contains precise but intermittent recordings. This is a closely related issue, however differs in the fact that when handling aggregated data, we do not have any precise times to work with.

We now outline the two methods against which we will compare our novel MC-EM approach.

\subsection{Hawkes INAR($p$) Approximation}\label{inar_section}

It is shown in \cite{kirchner_hawkes_2016,kirchner_estimation_2017} that the distribution of the bin-count sequence of Hawkes processes can be represented by an integer-valued autoregressive model, known as the INAR($p$) model, further details of which can be found in \cite{kirchner_hawkes_2016}. By representing the binned Hawkes counting process as an INAR($p$) process, a non-parametric estimator for kernel $g(u)$ is then formulated in terms of conditional least squares (CLS). 

Let $\Delta > 0$ be the bin width, the univariate Hawkes process bin-count sequence is denoted ${\mathbf{N}} = [N_1, \ldots, N_K]$, for $K=\lfloor T/\Delta \rfloor$ and $N_i$ denotes the counts in the $i^{\rm{th}}$ bin. 
Then, defining some support $\Delta<s<T$, the CLS-operator is used on the bin counts ${\mathbf{N}}$, with maximal lag $p= \lceil s/ \Delta \rceil$. 
Thus, 
\begin{align*}
{\mathbf{\hat{g}}}^{(\Delta,s)} \coloneqq \frac{1}{\Delta} {\mathbf{Y}} {\mathbf{Z}}^\top \left( {\mathbf{Z}} {\mathbf{Z}}^\top\right)^{-1},
\end{align*}
where the design matrix ${\mathbf{Z}}$ is given as 
\begin{equation*}
{\mathbf{Z}} = 
  \begin{bmatrix}
    N_p & N_{p+1} & \ldots & N_{k-1} \\
    N_{p-1} & N_p & \ldots & N_{k-2} \\
    \ldots & \ldots & \ldots & \ldots \\ 
    N_1 & N_2 & \ldots & N_{k-p} \\
    1 & 1 & \ldots & 1
  \end{bmatrix}
,
\end{equation*}
and ${\mathbf{Y}}$ is the lagged bin-count sequence, being $[ N_{p+1}, \ldots, N_k]$. Then the entries of ${\mathbf{\hat{g}}}^{(\Delta,s)}$, 
\begin{align*}
{\mathbf{\hat{g}}}^{(\Delta,s)} \coloneqq \left( \hat{g}_1^{(\Delta,s)}, \ldots, \hat{g}_p^{(\Delta,s)}, \hat{\nu}^{(\Delta,s)}  \right),
\end{align*}
are estimates for the excitation kernel at the corresponding time-points. Kernel parameters $\alpha$ and $\beta$ are then estimated by fitting an exponential function to these points.

Simulation studies examining the effect of bin width $\Delta$ and parameter $s$ are presented in \cite{kirchner_hawkes_2016,kirchner_nonparametric_2018}, where they determine $\Delta$ to have the greatest bearing on the quality of the estimates. There are however two points to note with this method. Firstly, CLS requires the inversion of $\mathbf{Z}\mathbf{Z}^\top$. In this case, this matrix contains the event counts per bin, and so it is possible to have cases where this matrix is singular, in particular when the counting process is very sparse. Secondly, as it is currently presented, this method does not constrain the parameter estimates to be those of a stationary Hawkes process. Therefore it is possible to yield infeasible estimates. 

\subsection{Binned Likelihood}
\label{sec:binnedll}
An alternative method, briefly mentioned in \cite{mark_network_2019} and developed here, considers sampling $\lambda^\ast$ at each discrete time-point $j\Delta$ ($j=1,...,K$ with $K=\lfloor T/\Delta\rfloor$), thus representing $\lambda^\ast$ as a piecewise constant function within each bin.
Letting $N_j$ be the number of events occurring in the sampling interval $((j-1)\Delta, j\Delta]$ and using (\ref{log_like_formula1}) we have that the log-likelihood of the underlying Hawkes process is
\begin{align}\label{log_like_discrete1}
\log \mathcal{L}(\Theta ; \mathbf{N}) &= \sum_{m=1}^{p} \sum_{j=1}^{K} N_j \log \left( \Delta \lambda^\ast_m(j \Delta) \right) - \Delta \lambda_m^\ast (j \Delta),
\end{align}
where $\lambda^\ast(j \Delta) \equiv \lambda^\ast(j \Delta| \mathcal{H}_{j \Delta})$ and $\mathcal{H}_{j\Delta}$ denotes the history of the process until time $j\Delta$. 

The assumption of a piecewise constant CIF is equivalent to assuming $N_j \sim$ Poisson$\left(\Delta \lambda^\ast \left( (j-1)\Delta \right) \right)$. However, it is important to note that this assumption is not correct as it ignores the excitation within each bin and therefore will be biased, especially in cases where the intensity is high relative to the bin width, $\Delta$. Nevertheless it provides us with a simple approximation. To estimate the process parameters we maximize (\ref{log_like_discrete1}) with constraints ensuring stationarity, as expressed by (\ref{eqn:branching_ratio}). 
In the case of an exponential excitation kernel, explicitly this implies $\nu, \alpha > 0$ and $\alpha / \beta < 1$.

We will now propose an alternative method of parameter estimation which iteratively uses `legal' sets of continuous candidate time-points and therefore does not assume a piecewise constant CIF.

\section{Monte Carlo EM Algorithm for Aggregated Data}\label{sec:MCEM}

The EM algorithm \cite{dempster_maximum_1977} is an iterative method for the computation of the maximizer of a likelihood. The idea of this algorithm is to augment the observed data by a latent quantity \cite{wei_monte_1990}. 
In the case considered here, the observed data are the event counts per unit time. We denote this by ${\mathbf{N}} = [N_1, \ldots, N_K]$, where $N_j$ denotes the counts in the $j^{\rm{th}}$ bin ($j=1,...,K$). 
The latent data, denoted $\mathcal{T}$ are the unobserved, true event times which are rounded on recording and the set of parameters to be estimated is denoted $\Theta = \{v, \alpha, \beta\}$. The algorithm proceeds as follows: 
\begin{enumerate}
\item In the E (Expectation) step, we evaluate
\begin{align}
\label{eqn:exactQ} 
Q_{i+1}(\Theta, \Theta^i) = \int_{\mathbb{T}^\ast} \log (p (\Theta \mid {\mathbf{N}}, \mathcal{T})) p ( \mathcal{T} \mid {\mathbf{N}}, \Theta^i) \diff \mathcal{T},
\end{align}
where $\mathbb{T}^\ast$ denotes the sample space for $\mathcal{T}$. That is, we compute the expectation $Q_{i+1}(\Theta, \Theta^i)$ of the log-posterior $\log (p (\Theta \mid {\mathbf{N}}, \mathcal{T}))$ with respect to the conditional predictive distribution $p ( \mathcal{T} \mid {\mathbf{N}}, \Theta^i)$, where $\Theta^i$ is the current, $i^{\rm th}$ approximation. 
\item In the M (Maximization) step, we update the value of the conditional expectation with its maximizer $\Theta^{i+1}$.
\end{enumerate}

When (\ref{eqn:exactQ}) is analytically intractable we require Monte Carlo methods for numerical computation. This is known as MC-EM \cite{wei_monte_1990}. If we are able to sample $\mathcal{T}$ directly from $p ( \mathcal{T} \mid {\mathbf{N}}, \Theta^i)$, then we can approximate the integral in (\ref{eqn:exactQ}) with
\begin{align*}
\frac{1}{m} \sum_{k=1}^m \log (p (\Theta \mid {\mathbf{N}}, \mathcal{T}^{\ast (k)})),
\end{align*}
where $\mathcal{T}^{\ast (k)}$ is the $k^{\rm{th}}$ Monte Carlo sample of $\mathcal{T}$. However, no such sampling regime is possible in the Hawkes process setting. We therefore use importance sampling to simulate a {\it{legal}} proposal for $\mathcal{T}$ (that is, a set of event times that match the binned counts) from an alternative distribution $q(\mathcal{T} \mid {\mathbf{N}}, \Theta^i)$ which is simple to sample from (see Section \ref{sec:sampling_method} for details). Each of these proposals is then weighted depending on the probability it came from the desired distribution.
That is, given a set of $m$ samples $\mathcal{T}^{\ast (1)},\ldots,\mathcal{T}^{\ast (m)}$, we assign weights
\begin{align}\label{eqn:weights}
w_k = \frac{p(\mathcal{T}^{\ast (k)} \mid {\mathbf{N}}, \Theta^i)}{q(\mathcal{T}^{\ast (k)} \mid {\mathbf{N}}, \Theta^i)},
\end{align}
and approximate (\ref{eqn:exactQ}) with
\begin{align}
\label{eqn:impQ}
Q_{i+1}(\Theta, \Theta^{i}) = \frac{\sum_{k=1}^m w_k \log (p (\Theta \mid {\mathbf{N}}, \mathcal{T}^{\ast (k)}))}{ \sum_{k=1}^m w_k  }.
\end{align}
We note that the numerator of (\ref{eqn:weights}) can be expressed as
\begin{equation*}
p\left(\mathcal{T}^{\ast (k)} \mid {\mathbf{N}, \Theta^i} \right) \propto p\left(\mathcal{T}^{\ast (k)}\mid \Theta^i \right) \prod_{j=1}^K \mathds{1}_{ {N_j} } \left( \sum_{i=1}^n \mathds{1}_{ [b_j^-,b_j^+) }\left(\mathcal{T}_i^{\ast (k)}\right) \right),
\end{equation*}
where  $\sum_{i=1}^n \mathds{1}_{ [b_i^-,b_i^+) }\left(\mathcal{T}_i^{\ast (k)}\right)$ is the number of event times in $\mathcal{T}^{\ast (k)}$ lying in bin $j$, and $\mathds{1}_{B}(y)$ is 1 if $y$ is in the set $B$ and 0 otherwise. Therefore, if only proposing legal event times, we have that 
$$
p \left( \mathcal{T}^{\ast (k)} \mid {\mathbf{N}, \Theta^i} \right) \propto p \left( \mathcal{T}^{\ast (k)} \mid \Theta^i \right),
$$
where, $\log \left( p(\mathcal{T}^{\ast (k)}\mid \Theta) \right)$ is given by (\ref{log_like_formula1}). 

However, the question remains of how to best sample the latent times. Here, the density we would ideally like to sample from is that of the missing event times given the bin counts and the model parameters $p(\mathcal{T} \mid  {\mathbf{N}}, \Theta^i)$. Therefore for this method to be most efficient and to ensure meaningful weights, the alternative distribution $q(\mathcal{T} \mid {\mathbf{N}}, \Theta^i)$ should be as close as possible to the true distribution of the time-stamps.
\subsection{Sampling Method}\label{sec:sampling_method}
It is possible to uniformly redistribute the events across a bin in order to generate a legal set of event times. However, especially for Hawkes processes with high activity, this is not optimal as it leads to weights that are too small to compute (\ref{eqn:impQ}). Therefore, we propose an alternative method  which samples from a distribution $q(\mathcal{T} \mid  {\mathbf{N}}, \Theta)$ that more closely matches $p(\mathcal{T} \mid  {\mathbf{N}}, \Theta)$. Here, this is developed for the exponential kernel but it is easily extendible to other kernels (see Appendices \ref{ap:pl} and \ref{ap:rec} for details regarding power-law and rectangular kernels, respectively).


Without loss of generality, consider having already simulated until time-point $t_n, n \in \{1,\ldots,N_T\}$. Then, suppose we know that there are $m$ time-points in the next non-empty bin, being $t_{n+1}, \ldots, t_{n+m}$. The joint probability of these events can be expressed using factorization. That is
\begin{align*}
&f^\ast_{T_{n+1},\ldots,T_{n+m}}(t_{n+1},\ldots,t_{n+m}) = \prod_{i=1}^{m} f^\ast_{T_{n+i}}(t_{n+i}),\\
&:= \prod_{i=1}^{m} f_{T_{n+i}}(t_{n+i} \mid \mathcal{H}_{t_{n+i}}) ,\\
&= \prod_{i=1}^{m} \lambda^\ast(t_{n+i}) \exp \left( -\int_{t_{n+i-1}}^{t_{n+i}} \lambda^\ast (u) \diff u \right).
\end{align*}
For brevity, we refer to $f^\ast_{T_{n+1},\ldots,T_{n+m}}(t_{n+1},\ldots,t_{n+m})$ as $f^\ast_{T_{n+1:n+m}}(t_{n+1:n+m})$.
Note that for the simplest case of an exponential kernel, we can express this as 
\begin{align*}
&\prod_{i=1}^{m} \left( \nu + \sum_{j=1}^{n+i-1} \alpha \exp (-\beta (t_{n+i} - t_j)) \right) \exp \left( -\int_{t_{n+i-1}}^{t_{n+i}} \nu + \sum_{j=1}^{n+i-1} \alpha \exp (-\beta (u - t_j)) \diff u \right),\\
&= \prod_{i=1}^{m} \left( \nu + \alpha A(n+i) \right) \exp(-\nu (t_{n+m} - t_n)) \exp \left( \sum_{i=1}^{m} \sum_{j=1}^{n+i-1} \frac{\alpha}{\beta} \left( e^{-\beta (t_{n+i}-t_j)} - e^{-\beta (t_{n+i-1}-t_j)} \right) \right),
\end{align*}
where
$$
A(n+i) = \sum_{j=1}^{n+i-1} \exp(-\beta (t_{n+i} - t_j)).
$$
As we wish to simulate possible realizations of the events given observed counts, we should account for the the fact that each time-point is known to have occurred within a given interval. That is, we account for the observed interval range $[b^-,b^+]$ for events in a given bin by considering the truncated joint density.  
We require that $b^- < t_{n+1} < t_{n+2} < \ldots < t_{n+m} < b^+$. Therefore, we can express the conditional CDF over this region as
\begin{align*}
\int_{b^-}^{b^+} \cdots \int_{b^-}^{t_{n+2}} f^\ast_{T_{n+1:n+m}}(t_{n+1:n+m}) \diff t_{n+1} \ldots \diff t_{n+m}.
\end{align*}
Even in the simplest case of an exponential decay kernel, this appears intractable due to the form of the conditional intensity function for a Hawkes process. Therefore we truncate the PDF by considering the joint CDF.
As with the joint PDF, we can use factorization to express the joint CDF as
\begin{align*}
&F^\ast_{T_{n+1},\ldots,T_{n+m}}(t_{n+1},\ldots,t_{n+m}) = 
\prod_{i=1}^{m} F^\ast_{T_{n+i}}(t_{n+i}),\\
&:= F_{T_{n+m}}(t_{n+m} \mid \mathcal{H}_{t_{n+m}})\ldots F^\ast_{T_{n+1}}(t_{n+1} \mid \mathcal{H}_{t_{n+1}}),
\end{align*}
where we can use the form given in (\ref{eqn:cond_cdf_pdf}) for the CDF of each successive time-point given the history of the process. That is, the joint CDF of time-points $t_{n+1}, \ldots, t_{n+m}$ is given by

\begin{align*}
\prod_{i=1}^{m} F^\ast_{T_{n+i}}(t_{n+i})
&= \prod_{i=1}^{m}
\Biggl( 1 - \exp\left( - \int_{t_{n+i-1}}^{t_{n+i}} \lambda^\ast(u) \diff u\right) \Biggr).
\end{align*}
In the case of an exponential decay kernel, this is
\begin{align*}
\prod_{i=1}^{m} \Biggl( 1 - \exp \Biggl( - &\int_{t_{n+i-1}}^{t_{n+i}} \nu  + \sum_{j=1}^{n+i-1} \alpha \cdot \exp(-\beta (u-t_j)) \diff u \Biggr) \Biggr).
\end{align*}
Thus the joint truncated PDF of $m$ time-points given the history, $t_1, \ldots t_n$ can be expressed as  
\begin{align}\label{truncated_pdf}
\frac{f^\ast_{T_{n+1},\ldots,T_{n+m}}({t}_{n+1},\ldots,{t}_{n+m})}{\kappa},
\end{align}
where
\begin{align*} 
\kappa = F^\ast_{T_{n+1},\ldots,T_{n+m}}({t}_{n+1},{t}_{n+2},\ldots,{t}_{n+m-1},b^+) - F^\ast_{T_{n+1},\ldots,T_{n+m}}(b^-,{t}_{n+2}\ldots,{t}_{n+m-1},{t}_{n+m}).
\end{align*}
The set of $m$ proposed time-stamps for the given bin $\{\tilde{t}_{n+1}, \ldots, \tilde{t}_{n+m}\}$ are those that maximize (\ref{truncated_pdf}). 


In this way we can sequentially simulate a continuous version of the observed aggregate Hawkes process by progressively handling each bin such that we jointly maximize this likelihood.


\section{Simulation Studies} 

\begin{figure}[!htb]
\centering
\includegraphics[scale=0.4]{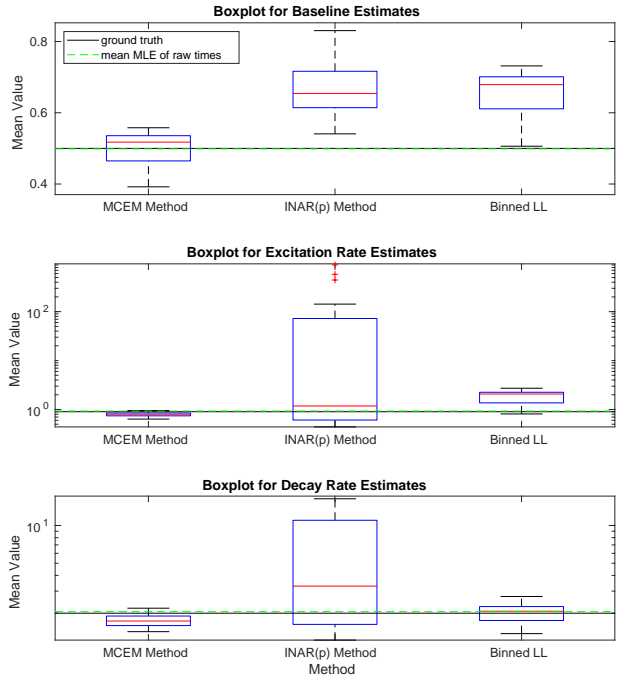}
\caption{Parameter set: $[\nu,\alpha,\beta] = [0.5,0.9,2.0]$, $\Delta = 1$. A log-scale is used for the excitation and decay parameter estimate boxplots to show extreme outliers.}
\label{fig:1}
\end{figure}

\begin{figure}[!htb]
\centering
\includegraphics[scale=0.4]{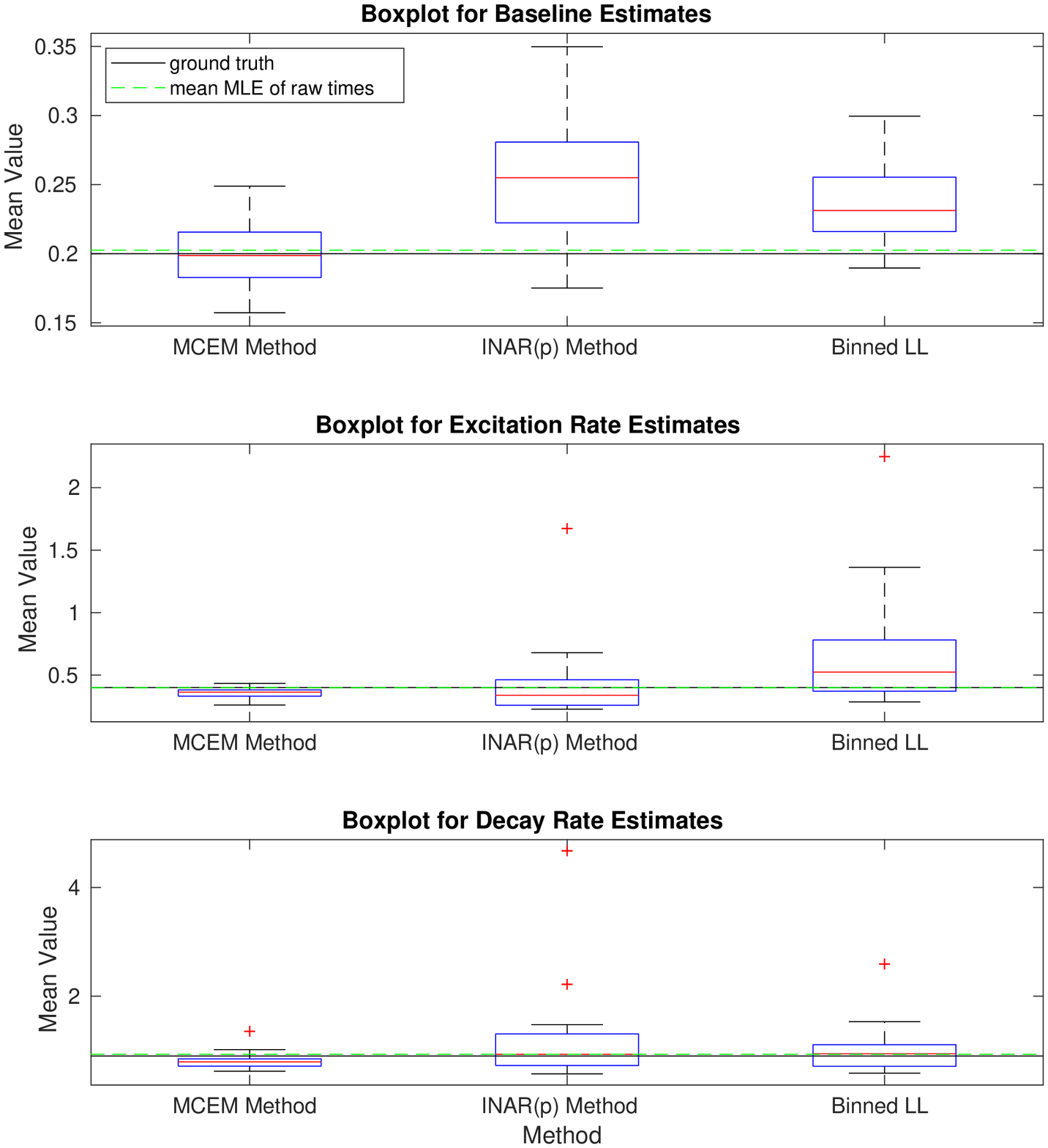}
\caption{Parameter set: $[\nu,\alpha,\beta] = [0.2,0.4,0.9]$, $\Delta = 1$.}
\label{fig:2}
\end{figure}

\begin{figure}[!htb]
\centering
\includegraphics[scale=0.4]{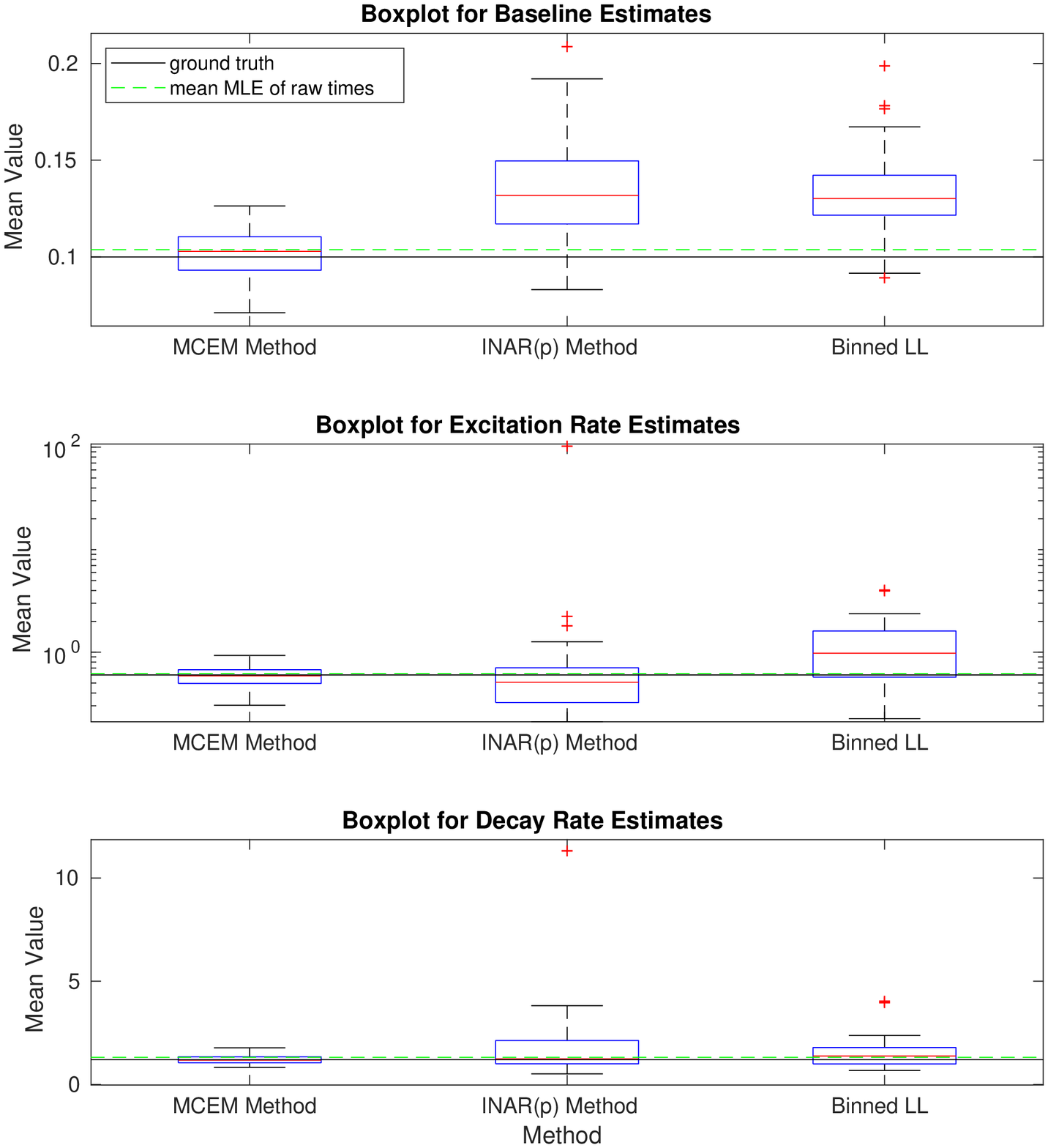}
\caption{Parameter set: $[\nu,\alpha,\beta] = [0.1,0.6,1.2]$, $\Delta = 1$. A log-scale is used for the excitation parameter estimate boxplot to show extreme outliers.}
\label{fig:3}
\end{figure}

\begin{figure}[!htb]
\centering
\includegraphics[scale=0.4]{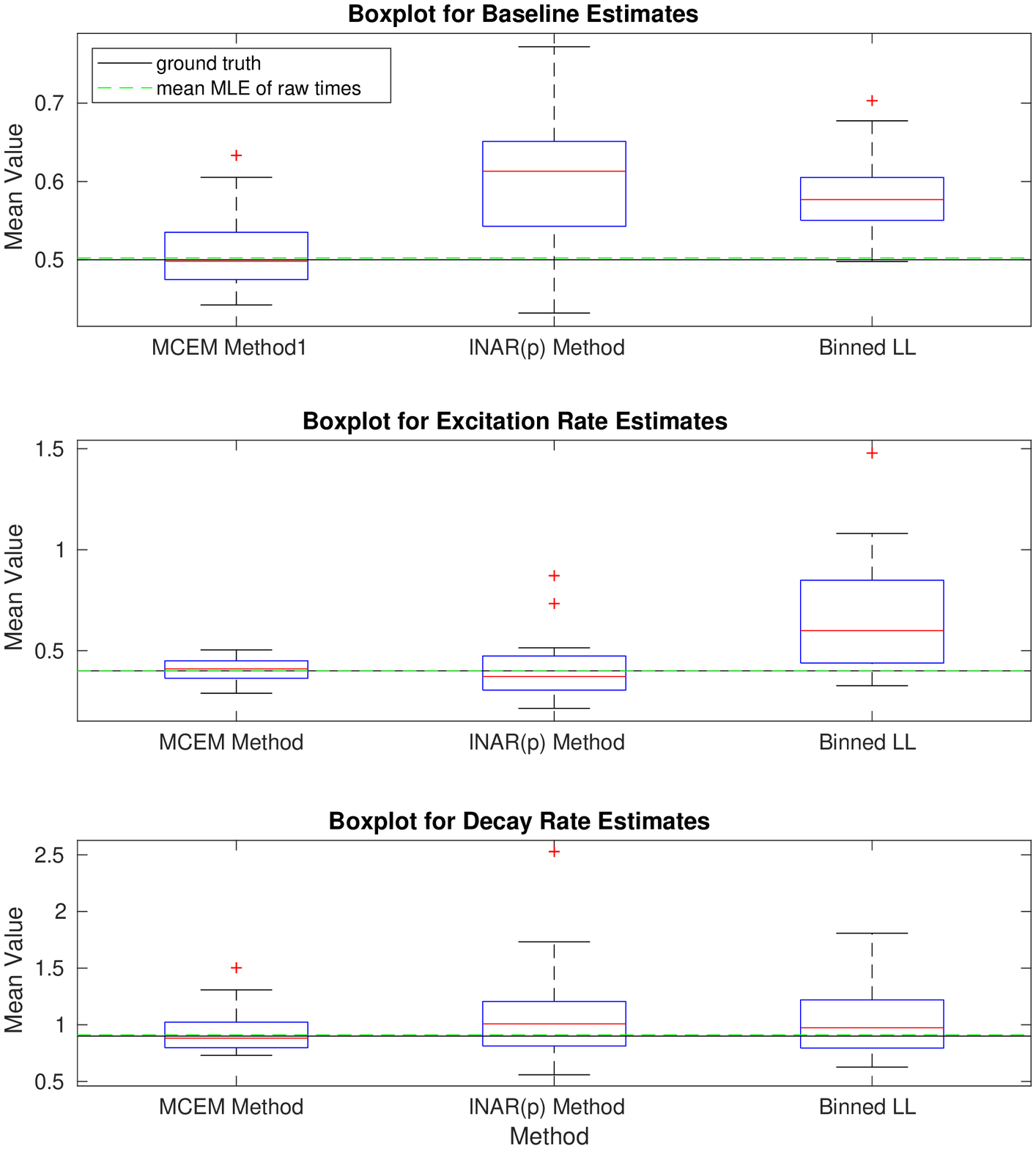}
\caption{Parameter set: $[\nu,\alpha,\beta] = [0.5,0.4,0.9]$, $\Delta = 1$.}
\label{fig:4}
\end{figure}

\begin{figure}[!htb]
\centering
\includegraphics[scale=0.4]{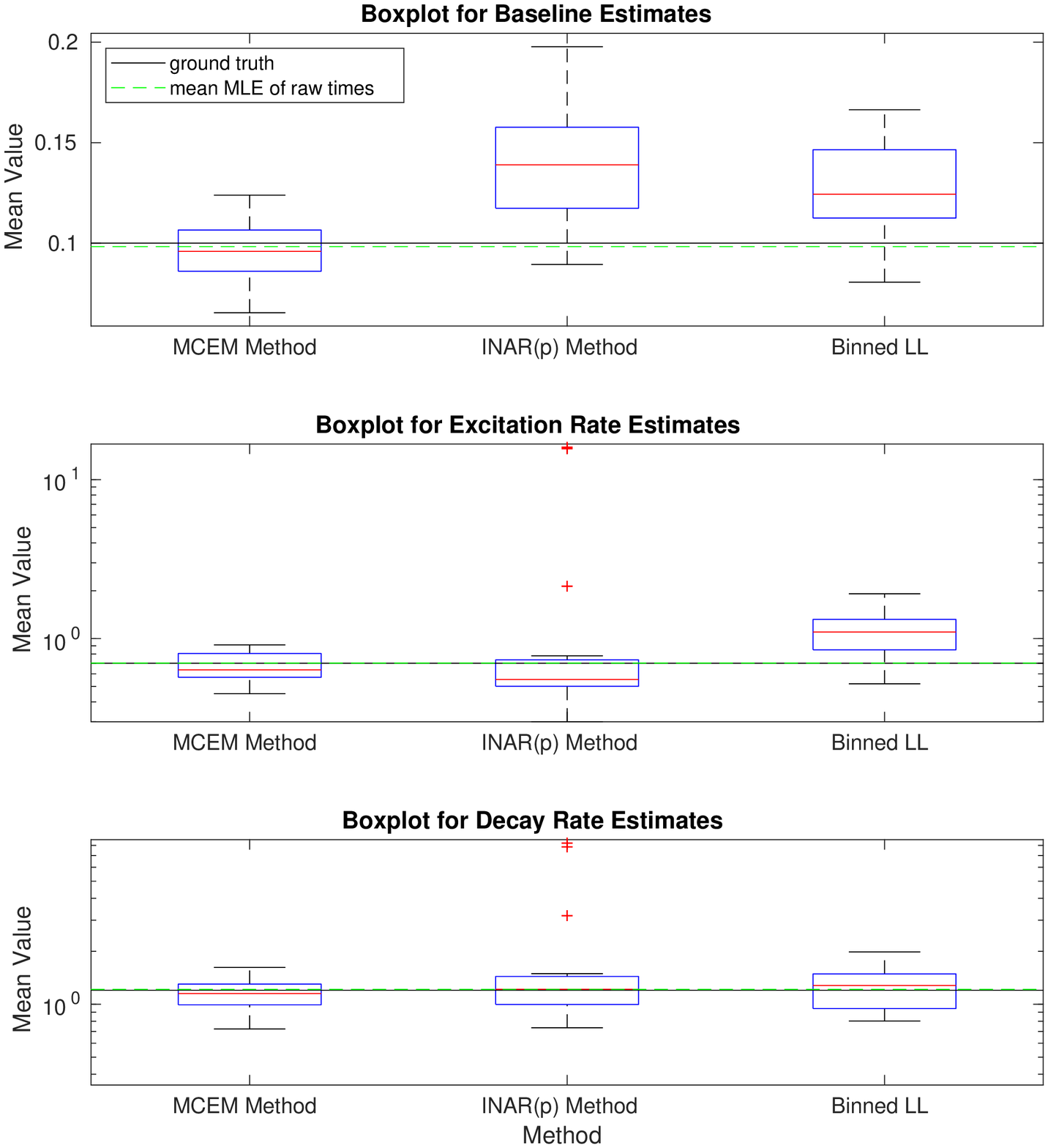}
\caption{Parameter set: $[\nu,\alpha,\beta] = [0.1,0.7,1.2]$, $\Delta = 1$. A log-scale is used for the excitation and decay parameter estimate boxplots to show extreme outliers.}
\label{fig:5}
\end{figure}

\begin{figure}[!htb]
\centering
\includegraphics[scale=0.4]{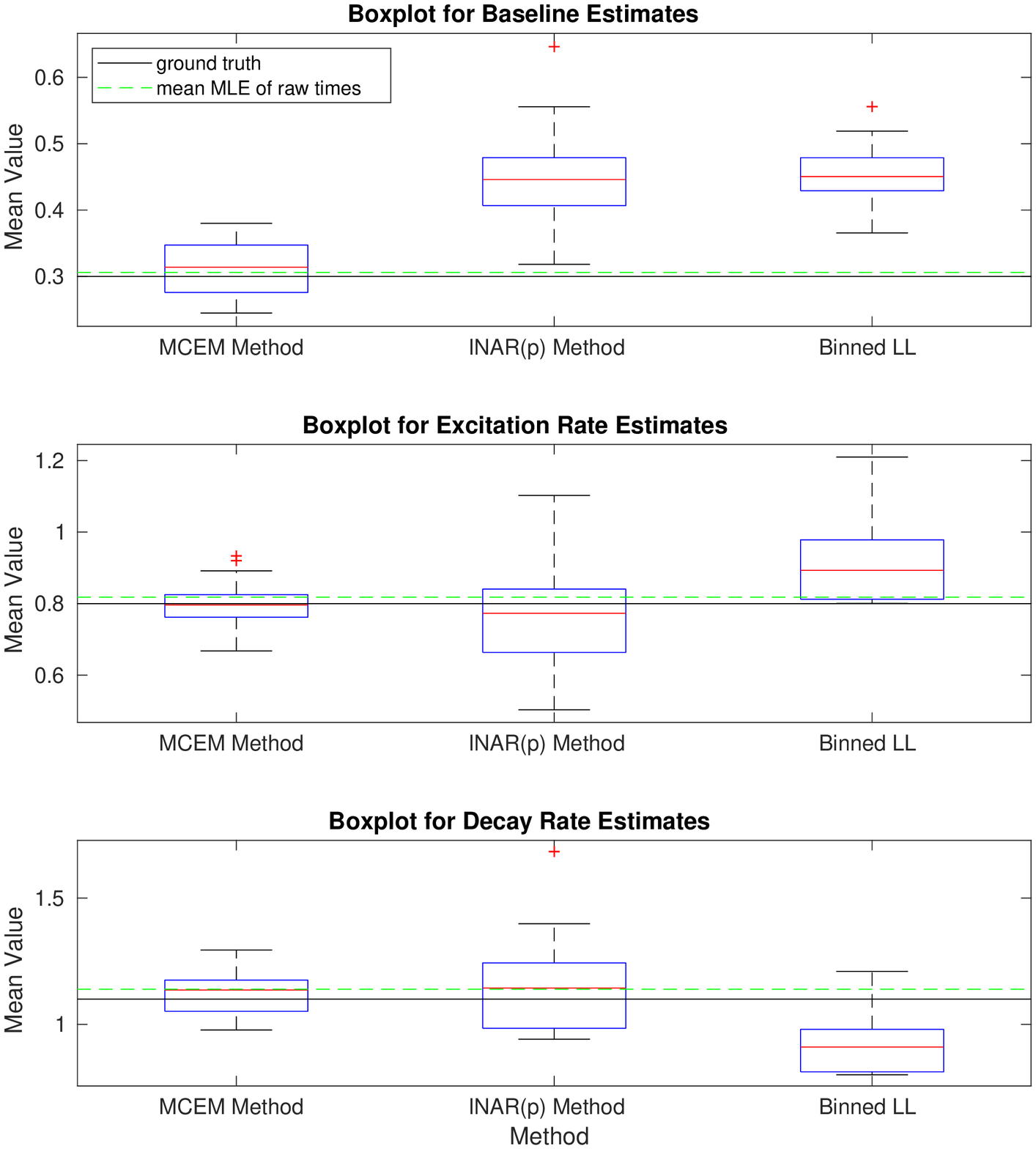}
\caption{Parameter set: $[\nu,\alpha,\beta] = [0.3,0.8,1.1]$, $\Delta = 1$.}
\label{fig:6}
\end{figure}

Given parameters $\nu, \alpha, \beta$ and some maximum simulation time $T$, we can simulate realizations of a Hawkes process. The generated events are those which form the latent space $\mathcal{T}$, and aggregating these to different $\Delta$ allows us to simulate the count data $\mathbf{N}$. We can then apply each of the three methods detailed: the binned log-likelihood, INAR($p$) approximation, and the MC-EM method. Figures \ref{fig:1}-\ref{fig:6} show boxplots for the estimates of each of $\nu, \alpha,$ and $\beta$ for 20 realizations of a Hawkes process with the ground truth parameters specified. The mean value of each parameter estimate is presented on the vertical axis. We clearly see that the INAR($p$) approximation method can yield highly variable results. In Figure \ref{fig:1}, the boxplots for both the excitation rate, $\alpha$, and the decay rate $\beta$ have been presented on a log-scale in order to show the results on one axis. Clearly, the INAR($p$) method has resulted in outliers that are factors of 10 away from the ground truth. The remaining figures all likewise show the MC-EM method to perform better than either of the two alternative approaches considered for a range of different parameter sets. In particular, again in Figure \ref{fig:5} we note that the excitation rate estimates have been presented on a log-scale due to the extreme outliers in the INAR($p$) method. The binned log-likelihood method, whilst suffering less from extreme outliers, performs less well than the MC-EM method in all cases.

\begin{figure*}[!htb]
\centering
\includegraphics[scale=0.4]{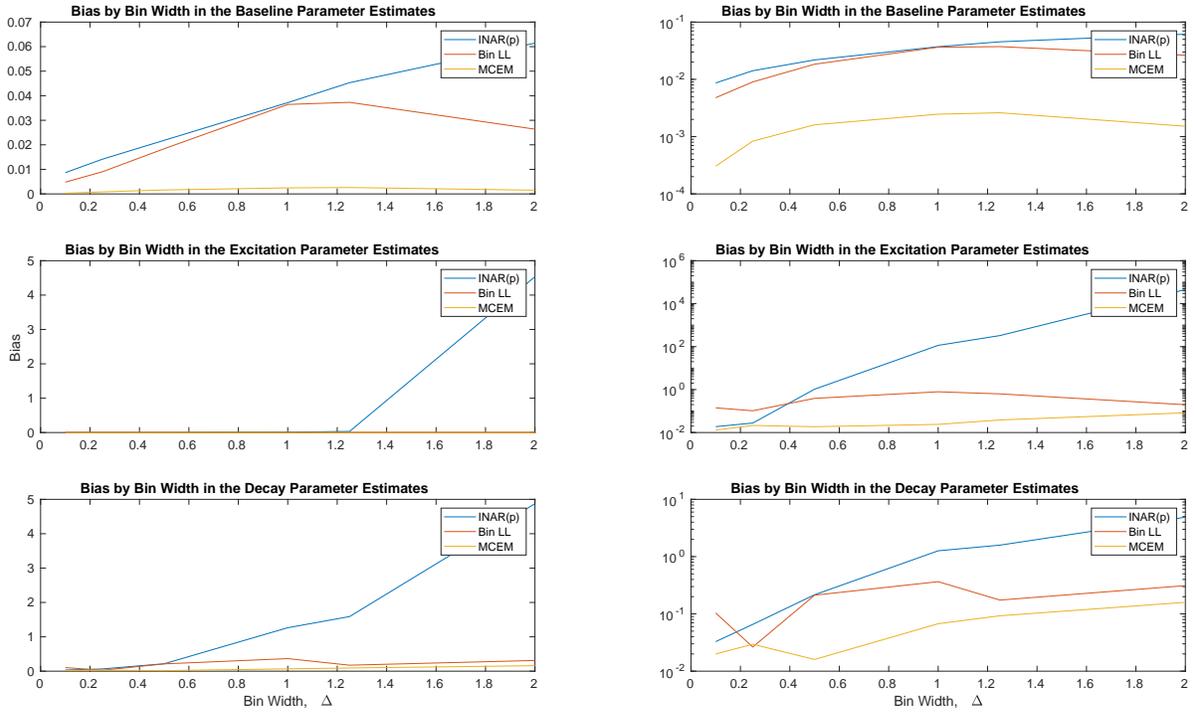}
\caption{Parameter set: $[\nu,\alpha,\beta] = [0.1,0.7,1.2]$, $\Delta = [0.1,0.25,0.5,1,1.25,2]$. Left figure shows the results on a linear scale, whilst the right shows a log scale.}
\label{fig:9}
\end{figure*}

In Figure \ref{fig:9} we also consider the bias across different levels of aggregation. That is, for each of the realizations of a self-exciting process for a given parameter set, we can aggregate the data to different levels and compare the bias in the parameter estimates. The right hand plot in Figure \ref{fig:9} presents the bias on a log-scale. It is evident that the INAR($p$) method is more biased for larger bin widths. The binned log-likelihood performs better, however, still not as well as the MC-EM method which most consistently exhibits a low bias. 

\FloatBarrier
\section{Conclusion}

Here we presented a new technique for handling aggregated data using an MC-EM algorithm. By sampling from a distribution close to that of the latent event times given the observed times and current parameter estimate, we have proposed a surrogate legal set of candidate time-points. This allows estimation of parameters using methods for continuous time-points. We further compared this to the INAR($p$) approximation proposed in \cite{kirchner_hawkes_2016} and a binned log-likelihood method which assumes a piecewise constant CIF within each interval. For the parameter sets considered, the MC-EM method has appeared to out-perform both alternatives. The MC-EM approach also provides us with additional flexibility in that the level of aggregation does not need to be consistent across the dataset. That is, provided the interval bounds are known, $\Delta$ can vary. The issues arising from aggregated data could also be handled via a MCMC (MCMH) algorithm and exploring this is the subject of future work.

\begin{appendices}
\section{Power-Law Kernel}
\label{ap:pl}
The proposed MC-EM method can still be applied if intending to consider a regularized power-law kernel of the form 
$$
g(u)=\frac{\alpha \beta}{(1+\beta u)^{1+c}} \mathds{1}_{u \in \mathbb{R}_+}(u),
$$
for $\alpha,\beta > 0$. In this case, to ensure a stationary Hawkes process, we have that
\begin{align*}
\gamma &= \int_0^{\infty}   \frac{\alpha \beta}{(1 + \beta u)^{1+c}} \diff u < 1,\\
&\implies \left[ - \frac{\alpha}{c} (1  +\beta u)^{-c} \right]_0^{\infty} < 1, \\
&\implies \frac{\alpha}{c} < 1.
\end{align*}
Therefore the stationarity condition is met for $\alpha < c$ \cite{bacry_hawkes_2015}.

We also need to consider the proposed univariate sampling method and thus the complete log-likelihood of sampled time-points. Both of these points fundamentally rely on expressing the conditional PDF and CDF. Firstly, for the sampling method introduced in Section \ref{sec:sampling_method}, we now have that 
\begin{align*}
&f^\ast_{T_{n+1},\ldots,T_{n+m}}(t_{n+1},\ldots,t_{n+m})
= \prod_{i=1}^{m} \lambda^\ast(t_{n+i}) \exp \left( -\int_{t_{n+i-1}}^{t_{n+i}} \lambda^\ast (u) \diff u \right), \\
&= \prod_{i=1}^{m} \left( \nu + \sum_{j=1}^{n+i-1} \frac{\alpha \beta}{(1 + \beta (t_{n+i}-t_{j}))^{1+c}} \right) \exp \left( -\int_{t_{n+i-1}}^{t_{n+i}} \nu + \sum_{j=1}^{n+i-1} \frac{\alpha \beta}{(1 + \beta (u-t_{j}))^{1+c}} \diff u \right),\\
&= \prod_{i=1}^{m} \left( \nu + \sum_{j=1}^{n+i-1} \frac{\alpha \beta}{(1 + \beta (t_{n+i}-t_{j}))^{1+c}} \right) \exp(-\nu (t_{n+m} - t_n)) \exp \Biggl( \sum_{i=1}^{m} \sum_{j}^{n+i-1} \frac{\alpha}{c} \cdot \Biggr. \\
&\Biggl. \left\{ (1 + \beta (t_{n+i}-t_{j}))^{-c} - (1 + \beta (t_{n+i-1}-t_{j}))^{-c} \right\} \Biggr).
\end{align*}

Similarly, the joint conditional CDF is given by 
\begin{align*}
&F^\ast_{T_{n+1},\ldots,T_{n+m}}(t_{n+1},\ldots,t_{n+m}) = \prod_{i=1}^{m} F^\ast_{T_{n+i}}(t_{n+i}), \\
&= \prod_{i=1}^{m}
\left( 1 - \exp\left( - \int_{t_{n+i-1}}^{t_{n+i}} \lambda^\ast(u) \diff u\right) \right).
\end{align*}
In the case of regularized power-law kernel, this is
\begin{align*}
\prod_{i=1}^{m} \Biggl( 1 - \exp \Biggl( - &\int_{t_{n+i-1}}^{t_{n+i}} \nu + \sum_{j=1}^{n+i-1} \frac{\alpha \beta}{(1 + \beta (t_{n+i}-t_{j}))^{1+c}} \diff u \Biggr) \Biggr).
\end{align*}
Then, (\ref{truncated_pdf}) gives the form for the truncated PDF, as previously. All that remains is to adjust the log-likelihood for the CIF with regularized power-law function when implementing the MC-EM algorithm. Using (\ref{log_like_formula1}) that is, 
\begin{align*}
&\log \mathcal{L}(\Theta ; \mathcal{T}) = \sum_{i=1}^{N_T} \log \left( \nu + \sum_{j=1}^{n+i-1} \frac{\alpha \beta}{(1 + \beta (t_{n+i}-t_{j}))^{1+c}}  \right) - \int^{T}_{0} \nu + \sum_{j=1}^{n+i-1} \frac{\alpha \beta}{(1 + \beta (u-t_{j}))^{1+c}} \diff u.
\end{align*}

\section{Rectangular Kernel}
\label{ap:rec}
We can also consider a rectangular kernel of the form 
\begin{align*}
g(u) &= \frac{n}{\beta-\alpha} \mathds{1}_{[\alpha,\beta]} (u), \\
&\equiv \frac{n}{\beta-\alpha} \mathds{1}_{[0,\beta-\alpha]} (u-\alpha),
\end{align*}
for $\alpha, \beta > 0$. In this case, stationarity holds if
$$
\int_{\alpha}^{\beta} \frac{n}{\beta-\alpha} \diff u = n < 1. 
$$
Note that $\alpha$ here represents a small shift of the excitation effect. Therefore, if  $\alpha=0$, there is an increase in the process intensity immediately after an arbitrary event. In the case of a rectangular kernel, 
\begin{align*}
\lambda^\ast (t) &= \nu + \sum_{t_i < t} \frac{n}{\beta-\alpha} \mathds{1}_{[\alpha,\beta]} (t - t_i),\\
&= \nu + \frac{n}{\beta-\alpha} \sum_{t_i < t} \mathds{1}_{[\alpha,\beta]} (t - t_i).
\end{align*}
The remaining equations follow as previously by substituting the above CIF. 

\section{The MC-EM algorithm}
\begin{algorithm}
\caption{Sampling from alternative distribution $q(\mathcal{T}^{\ast (j)} \mid {\mathbf{N}}, \Theta^i)$}
\label{pseudoSampling}
\begin{algorithmic}[1]
\Function{Sample from $q$}{${\mathbf{N}}, \Theta^i$}
\State $\mathcal{T}^{\ast (j)} \gets \emptyset$
\For{$i = 1$ to $\lvert \{ N_k \in {\mathbf{N}}: N_k \neq 0 \} \rvert$}
	\State $\{\tilde{t}_{1}, \ldots, \tilde{t}_{m}\} \gets \rm{set \; which\; maximizes}$ (\ref{truncated_pdf}) for $m = N_i$
	\State $\mathcal{T}^{\ast (j)} \gets \mathcal{T}^{\ast (j)} \cup \{\tilde{t}_{1}, \ldots, \tilde{t}_{m}\}$
	
\EndFor
\EndFunction 
\end{algorithmic}
\end{algorithm}

\begin{algorithm}
\caption{MC-EM}
\label{pseudoMCEM}
\begin{algorithmic}[1]
\Function{MCEM}{${\mathbf{N}},m,\epsilon$}
\State $\Theta^1 \gets \rm{sort}(\rm{Unif}(1,3))$ 
\State $i \gets 1$
\While{tolerance $> \epsilon$}
	\For{$j = 1$ to $m$}
		\State $\mathcal{T}^{\ast(j)} \sim q(\mathcal{T} \mid {\mathbf{N}}, \Theta^i)$, provided in Algorithm \ref{pseudoSampling}
		\State $w_j \gets p(\mathcal{T}^{\ast (j)} \mid \Theta^i)/q(\mathcal{T}^{\ast (j)} \mid {\mathbf{N}}, \Theta^i)$
	\EndFor
    \State $Q_{i+1}(\Theta, \Theta^{i}) \gets \sum_{k=1}^m w_k \log (p (\Theta \mid {\mathbf{N}}, \mathcal{T}^{\ast (k)})) / \sum_{k=1}^m w_k$
	\State $\Theta^{i+1} \gets \argmax_{\Theta, \gamma<1} Q_{i+1}(\Theta, \Theta^{i}) $	
	\State tolerance $\gets \rm{norm}(\Theta^{i+1} - \Theta^{i}) $
	\State $i \gets i + 1$
\EndWhile
\State \textbf{return} $\{\Theta^i\}$ \Comment{Set of parameter estimates}
\EndFunction 
\end{algorithmic}
\end{algorithm}

\end{appendices}
\FloatBarrier

\section*{Acknowledgment}
Leigh Shlomovich is funded by an Industrial Strategy Engineering and Physical Sciences Research Council Scholarship and Lekha Patel is funded by an Imperial College President's Scholarship.


%

%
%

\bibliographystyle{abbrv}  
\bibliography{arxiv_mcem}

\end{document}